\documentclass[aps,prl,reprint,showpacs]{revtex4-1}

\usepackage{graphicx}
\usepackage{subfigure}
\usepackage{bbm}
\usepackage{amsmath}
\begin{document}

\title{Coherence and Entanglement Preservation \\ of Frequency-Converted Heralded Single Photons}

\author{Andreas Lenhard}
\author{Matthias Bock}
\author{Christoph Becher}
\affiliation{Quantenoptik, Universit\"at des Saarlandes, Campus E2.6, 66123 Saarbr\"ucken, Germany}

\author{Jos\'e Brito}
\author{J\"urgen Eschner}
\affiliation{Quantenphotonik, Universit\"at des Saarlandes, Campus E2.6, 66123 Saarbr\"ucken, Germany}

\begin{abstract}
We report on quantum frequency conversion of near-infrared photons from a wavelength of 854~nm to the telecommunication O-band at 1310~nm with 8~\% overall conversion efficiency. Entangled photon pairs at 854~nm are generated via type-II spontaneous parametric down conversion. One photon is mixed with a strong pump field in a nonlinear ridge waveguide for its conversion to 1310~nm. We demonstrate preservation of first and second order coherence of the photons in the conversion process. Based on this we infer the coherence function of the two-photon state and compare it with the actual measured one. This measurement demonstrates preservation of time-energy entanglement of the pair. With 88~\% visibility we violate a Bell inequality. 
\end{abstract}

\maketitle

\section{Introduction}

Long-haul quantum communication requires the transfer of entanglement between remote locations with the help of quantum repeaters \cite{Bri98}. Such an architecture relies on photonic qubits to establish entanglement between neighboring atomic nodes \cite{Dua01, Sim07}. Thus the photons have to be resonant with an atomic transition of the node. As such resonances are usually found in the near infrared (NIR), the photons will suffer high attenuation in optical fibers. The losses in fibers are minimal at the so called telecom bands between 1260-1650~nm. Quantum frequency conversion (QFC) is a promising technique to bridge the gap between NIR and telecom bands by transducing the wavelength of a photonic state while preserving its other properties. QFC has already been used to transfer single photons between telecommunication wavelengths and NIR either by up-conversion \cite{Rak10, Ate12} or down-conversion \cite{Zas12} schemes. Furthermore, the entanglement of a photon with an atomic system \cite{Gre12} as well as with another photon \cite{Tan05, Iku11, Ram12} is preserved in the conversion process. Hence, along with the development of asymmetric (i.e., non-degenerate) photon pair sources \cite{Fek13, Len15}, QFC is a promising technique for interfacing quantum memories with telecom wavelengths \cite{Alb14}. Beyond the storage of qubits, single atoms allow one to implement protocols for quantum information processing, as demonstrated with trapped Calcium ions \cite{Sch13}. With this species various interface operations have been demonstrated such as heralded absorption of single photons \cite{Pir11}, photonic entanglement in the absorption process \cite{Huw13}, bidirectional conversion between atomic and photonic qubits \cite{Stu12, Kur14, Cas15, Kur16} and interconnection of distant ions via photons \cite{Moe07, Sch13b, Huc15}. To implement Ca$^+$ based nodes into a future quantum network infrastructure we here demonstrate QFC between 854~nm, a transition wavelength in Ca$^+$, and the telecom O-band around 1310~nm. 

Time-energy or time-bin entanglement seem most feasible for transfer in long-haul fiber networks due to their high immunity to decoherence. Such qubits have already been used for quantum cryptography with unconditional security \cite{Tit00, Zha14}, as well as for entanglement transfer \cite{Cla11}. QFC is compatible with time-bin \cite{Iku11} or time-energy \cite{Tan05} entanglement, as earlier reports have investigated. Time-energy entanglement is an intrinsic property of photon pairs generated by spontaneous parametric down conversion (SPDC). A common way to prove the entanglement is a two-photon interference experiment in a Franson type setup \cite{Fra89}. Similar to single-photon interference effects the interference visibility depends on interferometer detunings, and Franson interference can thus be used to measure the coherence function of the two-photon state. Apart from such a characterization of fundamental quantum properties, Franson interference has been applied for long-range quantum communication \cite{Tit99} based on the time-energy entanglement of photon pair states. 

In this article we report on QFC of one photon of a near-infrared SPDC pair, tailored to be resonant with a transition in Ca$^+$ \cite{Pir11, Huw13}, to the telecom O-band. We investigate in detail the first- and second-order coherence of the photons before and after frequency conversion. Moreover we employ Franson interferometry to demonstrate the time-energy entanglement between the NIR photon and its telecom-converted partner and derive the two-photon coherence properties of the pair state. We present a model connecting the coherence properties of the two-photon state with that of the single photons. The model accounts for spectral filtering during the QFC process. With this detailed investigation we demonstrate that QFC is a versatile tool in quantum information processing and networking, interfacing atomic transitions with the telecom range and preserving initial quantum properties.

\section{Frequency Conversion}\label{sec:frequency_conversion}

\begin{figure*}[hbt]
	\centering
		\includegraphics[width=0.90\textwidth]{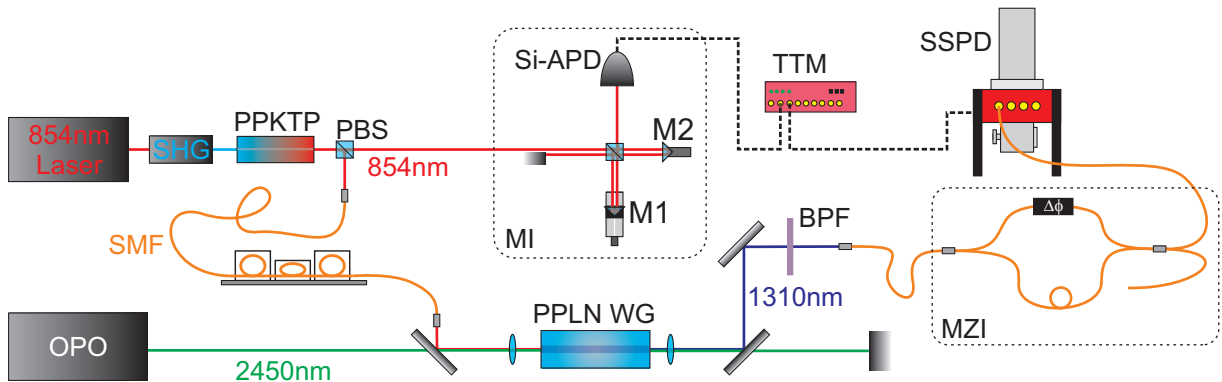}
	\caption{Experimental setup with SPDC source, frequency converter and interferometers for Franson interference. PBS: polarizing beam splitter; TTM: time tagging module; PPLN WG: periodically poled lithium niobate waveguide; BPF: band pass filters; PPKTP: periodically poled KTP nonlinear crystal; MI: Michelson interferometer; MZI: Mach-Zehnder interferometer.}
	\label{fig:FransonQFCSetup}
\end{figure*}

We use the second-order nonlinear process of difference frequency generation for quantum frequency conversion (QFC) between $\lambda_s=854$~nm (signal field) and the telecommunications O-band around $\lambda_i=1310$~nm (idler field). In this case the idler field is stimulated by a strong classical pump field at a wavelength around $\lambda_p=2453$~nm ($1/\lambda_i=1/\lambda_s-1/\lambda_p$). The pump field is generated by a home-built continuous-wave optical parametric oscillator (OPO). The OPO output wavelength is tunable between 2311 and 2870~nm with single mode, single frequency output power around 1~W. Hence we can tune the target wavelength of the conversion process over the entire telecom O-band. The pump power for the conversion process is controlled by a combination of a waveplate and a polarizer. For efficient down-conversion at the single photon level we use ridge waveguide structures (fabricated by NTT) based on a periodically poled lithium niobate core. The chip contains several wave\-guides with different poling periods for coarse tuning of the phase matching conditions, covering the whole telecom O-band; fine tuning of the phase matching is realized by the device temperature. 
Coupling both signal and pump field into the waveguide with good spatial mode overlap is mandatory for high conversion efficiency. The diameter of each beam was individually optimized before combining both on a dichroic mirror. After overlapping, both beams are focused to the waveguide with a single uncoated aspherical lens (Thorlabs A220TM). We reach a coupling efficiency, including transmission loss, of 84.2\% at 854~nm and 38\% at 2.5~$\mu$m. We suffer from absorption by the lens material resulting in only 72.6\% transmission at 2.5~$\mu$m. This can be improved by using ZnSe lenses which have high transmission over the whole wavelength range. Behind the waveguide the telecom light is collimated by another aspheric lens. Residual signal photons are separated off by a dichroic mirror. The coating of all following mirrors is optimized for telecom wavelengths, introducing considerable loss for the pump field. Finally, the telecom photons pass three interference band-pass filters (2x 50~nm FWHM, 1x 12~nm FWHM) with a total transmission of 37\%. This suppresses residual pump light and background photons induced by the pump field via Raman scattering \cite{Zas11}, second harmonic generation or higher order nonlinear processes. Then the photons are coupled to a standard telecom fiber (SMF28) with 80\% efficiency. Including these losses we measure an overall conversion efficiency of 8\% for the maximum available pump power. We use superconducting single photon detectors (SSPD, Single Quantum EOS X10) with 25\% detection efficiency for counting the telecom photons. The complete setup is sketched in Fig.~\ref{fig:FransonQFCSetup}.

Photons at 854~nm are generated through spontaneous parametric down conversion (SPDC) with a source described in more detail earlier \cite{Haa09, Pir09, Pir10}. We start with a diode laser system at 854~nm (Toptica DL-Pro), actively stabilized to an atomic transition ($3^2{\rm D}_{5/2} \leftrightarrow 4^2{\rm P}_{3/2}$ in $^{40}{\rm Ca}^+$). After second harmonic generation (SHG) to 427~nm, that light serves as pump for the SPDC process. We use a 2~cm crystal of bulk, periodically poled KTP (PPKTP) with type-II phase matching conditions to generate frequency-degenerate, polarization-entangled photon pairs at 854~nm. For the particular experiments reported here, the polarization entanglement is not necessary. Hence we separate the photons of each pair by a polarizing beam splitter. One photon serves as a herald while the other one is coupled to a single mode fiber and sent to the frequency converter setup. The converter is situated in a different lab, connected by 90~m of fiber.

The spectrum of the 854~nm photons is shown in the inset of Fig.~\ref{fig:854Messungen}. It was measured with a grating-based spectrometer of 1800 lines/mm. The spectrum has a width of 173~GHz (FWHM), slightly larger than the phase matching bandwidth of the QFC process (calculated 118~GHz FWHM). Hence the conversion efficiency will be reduced for the outer spectral components. The resulting spectrum of the converted photons will be the product of phase matching and original spectrum. As our spectrometer for the telecom wavelength range does not resolve the spectrum in sufficient detail, we gather more information about the spectral properties from a measurement of the first-order temporal coherence.

\section{First-order Coherence}\label{sec:first_order_coherence}

To measure the first-order coherence function we set up a Michelson-type interferometer. One mirror (M1) is mounted on a linear translation stage moved by a stepper motor and offers a coarse setting of the path length difference corresponding to a delay $\tau$ of up to 1.7~ns with 7~fs minimum step size. The mirror in the other arm (M2) is mounted on a piezo translator for fine movement. The input and output beam paths are coupled to single mode fibers. We use silver coated mirrors to allow operation over a wide wavelength range. The interferometer can be used for 854~nm as well as for 1310~nm by exchanging the fiber couplers and the 50/50 beam splitter and quick realignment.

\begin{figure}[htb]
	\centering
		\includegraphics[width=0.45\textwidth]{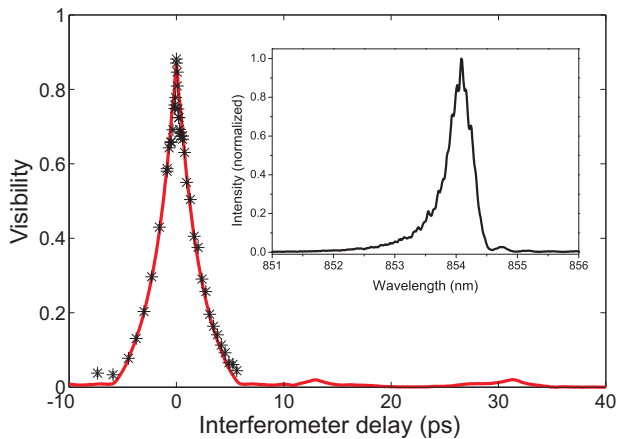}
	\caption{Measured visibility of the 854~nm single photons as a function of interferometer delay (black stars). The red solid line shows the calculation derived from the spectrum, shown in the inset.\label{fig:854Messungen}}
\end{figure}

We start with a measurement of the first-order coherence function of the original 854~nm SPDC photons. To this end, the delay is set via the translation stage to several values around the zero delay position. At each position the mirror M2 is scanned via the piezo translator and the count rate behind the interferometer is recorded with 100~ms integration time. The observed interference fringes are fitted by a sine function and their visibility as a function of the delay, $V(\tau)$,
is calculated from the fit. In a preparatory measurement with coherent laser light we found a maximum visibility for the measurement apparatus of $V_\textrm{max}=88$\% which we attribute to imperfect alignment and beam overlap in the interferometer. The measurement result for the 854~nm SPDC photons is shown in Fig.~\ref{fig:854Messungen}. We also achieve a maximum visibility of 88\% which is hence limited by the measurement device.

The first-order coherence function is defined as $g^{(1)}(\tau)=\frac{\langle E^*(t) E(t+\tau) \rangle}{\left[ \langle |E(t)|^2 \rangle \langle |E(t+\tau)|^2 \rangle \right]^{1/2}}$. This normalized function oscillates with the frequency of the optical field, and its envelope corresponds to the normalized visibility $V(\tau)/V_\textrm{max}$. Apart from the fast oscillating phase, the visibility function $V(\tau)$ thus contains all information on the first-order coherence. 
As we have additionally measured the spectrum of the photons with high resolution (see inset of Fig.~\ref{fig:854Messungen}), we can also derive the envelope of $g^{(1)}(\tau)$ via Fourier transformation of the power spectrum. The result is shown as solid line in Fig.~\ref{fig:854Messungen}, where only the peak height is fitted to the measured data. The shape of both curves overlap well, proving the consistency of the two measurements in the time and frequency domain. 

We estimate the coherence time $\tau_c$ by integrating the first-order coherence function:
\begin{equation}
	\tau_c = \int_{-\infty}^{\infty} d\tau \left|g^{(1)}(\tau)\right| 
	=  \int_{-\infty}^{\infty} d\tau V(\tau)/V_\textrm{max}
\end{equation}
With this method we find $\tau_{854}=3.7$~ps for the coherence time of the 854~nm photons.
In analogy we define the spectral bandwidth of the photons by integration of the normalized spectral intensity (of Fig.~\ref{fig:854Messungen}):
\begin{equation}
	\Delta\nu = \int_{-\infty}^{\infty} d\nu I(\nu)/\text{max}(I(\nu))
\end{equation}
This yields a robust value that allows us to compare bandwidths of spectra of dissimilar shapes, as reported in this paper. Here it results in $\Delta\nu_{854}=308$~GHz. With the help of these two quantities we now define the time-bandwidth-product as ${\it TBP} = \tau_c\cdot\Delta\nu$. For the original photons we find ${\it TBP}_{854}=1.14$.

\begin{figure}[tbh]
	\centering
		\includegraphics[width=0.45\textwidth]{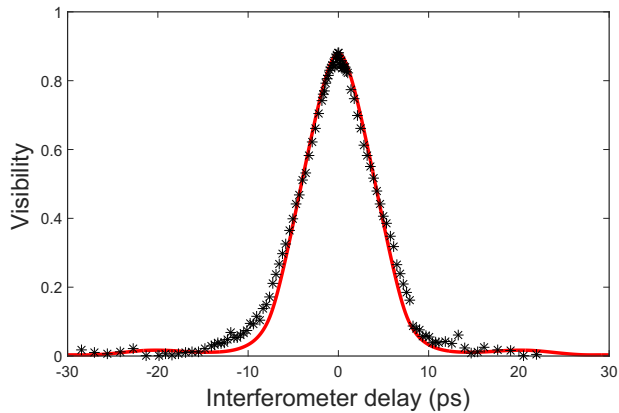}
	\caption{Measured visibility function of the converted 1310~nm telecom single photons (black stars). The solid line shows a calculation, as explained in the text.\label{fig:1310Messungen}}
\end{figure}

We repeat the measurement for the converted photons after exchanging the optics set of the interferometer as described before. The resulting visibility function of the telecom photons is shown in Fig.~\ref{fig:1310Messungen}.
In the same way as for the photons at the original wavelength we determine a coherence time of $\tau_{1310}=10.5$~ps which is significantly larger than before. This is due to the fact that the phase matching spectrum of the conversion process is narrower than the spectrum of the SPDC photons; the QFC process acts as a spectral filter. This results in a reduced bandwidth after the conversion process and hence a larger coherence time. In addition, we use a narrow-band pump-field with long coherence time which does not contribute to the bandwidth of the converted photons. We estimate the spectrum of the converted photons as the product of the original spectrum and the sinc$^2$-function of the QFC phase matching function ($\Delta\nu_{\text{sinc}^2}=131.4$~GHz). The resulting spectrum has a bandwidth of $\Delta\nu_{1310}=101.3$~GHz, resulting in ${\it TBP}_{1310}=1.06$, which is close to the original value. When we calculate the Fourier transform of this inferred spectrum we also obtain the expected shape of the coherence function, including the effect of the phase matching curve. The resulting visibility function is shown as a solid line in Fig.~\ref{fig:1310Messungen}, where only  the peak value is fitted to the measured data. The good agreement of its shape with the measured data confirms the validity of our model and shows that the QFC process preserves the coherence properties of the photons.

\section{Franson Interference}

Due to their generation process the SPDC photon pairs are time-energy entangled. Here we investigate to which degree this entanglement is preserved in the frequency conversion process. Entanglement preservation is expected, due to the fact that we use a continuous wave pump field with long coherence time for the conversion. We measured the frequency of the pump field with the help of a wavemeter (High Finesse WS6-200): the peak to peak fluctuations were below 200~MHz over 10~h of measurement time, and the Allan deviation of the frequency fluctuations was below 2~MHz for integration times below 100~s. Hence we affirm that the bandwidth of the pump field is much smaller than that of the signal photons and should not disturb the temporal properties of the photons, apart from introducing a constant and well-defined frequency shift. However, we already saw that the coherence function changes due to the finite width of the phase matching function. One common procedure to study the time-energy entanglement of photon pairs was proposed by Franson \cite{Fra89}. It involves sending both photons of the pair through individual interferometers which are equally unbalanced, and detecting the photon coincidences at the interferometer outputs. Interference fringes with visibility exceeding 0.5 are a signature for entanglement \cite{Fra91}. The experimental details will be discussed in the following. Such an experiment was first reported by Tanzilli at al.\ \cite{Tan05} for frequency up-conversion. Our experiment, in contrast, is based on frequency down-conversion, and furthermore we measure the coherence function of the photon pair state over a wide delay range. The results are compared with the theoretical expectations derived from the single-photon interference experiments.

To observe Franson interference two interferometers are necessary: for the 854~nm photons we use the Michelson interferometer (MI) described earlier. For the converted photons we set up another interferometer in Mach-Zehnder (MZI) geometry. As it is for telecom wavelengths we set it up from fiber components. It consists of two fiber beam splitters, a phase modulator (Phoenix photonics) and a delay fiber. All components are non-polarization-maintaining but we control the polarization state by fiber strain. In a preparatory measurement with a laser we found a maximum achievable visibility of 95~\%. To observe interference fringes in coincidence detection, we suppress first-order interference by choosing the delay much larger than the coherence time of $\tau_{1310}=10.5$~ps. On the other hand, the delay must be smaller than the coherence time of the pump laser generating the photon pairs (line width in the low MHz range). In our case the delay is limited to 1.7~ns by the size of the MI. We chose a delay of 1.14~ns, fixed by the length of the delay fiber in the MZI and fulfilling all requirements. The two photons of a pair are generated simultaneously within the pump coherence time and then travel either the short or the long path in their respective interferometers. To erase which-path information the delays in both interferometers have to be equal. We set the delay of the MI via the motorized mirror (M1) to match the one of the MZI. Then mirror M2 of the MI is scanned on the wavelength scale, and (time-shifted) coincidences between the 854~nm photons passing the MI (detected with a Perkin Elmer SPCM-AQR-14 Si-APD) and the converted telecom photons passing the MZI (detected with the SSPD) are recorded. 

\begin{figure}[htb]
	\centering
		\includegraphics[width=0.45\textwidth]{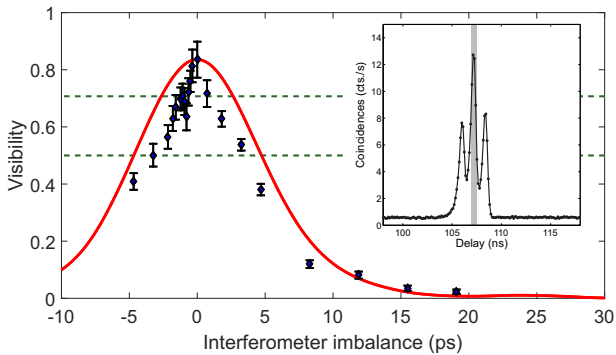}
	\caption{Two-photon coherence measurement. Main figure: measured interference fringe visibility (black diamonds) vs.\ delay imbalance between the interferometers. The red solid line shows the expected curve calculated from the first-order coherence functions and scaled by the maximum visiblity of the apparatus. The horizontal line at a visibility of 71~\% indicates the upper boundary given by the Bell inequality, and the line at 50~\% indicates the classical boundary. The inset shows a long-term average correlation function between the two detectors. The gray shaded area indicates the indistinguishable path combinations for which the visibility is determined. The coincidence data are background-corrected. \label{fig:FransonQFC_FranG1}}
\end{figure}

The measured coincidence function is shown as inset in Fig.~\ref{fig:FransonQFC_FranG1}. We observe three peaks due to the different path combinations. The dips between the peaks do not reach zero due to the timing jitter ($\sim 600$~ps) of our Si-APD single photon detector. The central peak contains the coincidence events from indistinguishable paths, i.e.\ where both photons travel either the long or the short path in their individual interferometers. Hence we cut out the coincidences from this central peak in the post processing, using a gate window of 512~ps as indicated by the shaded region. The visibility of the Franson interference vs.\ interferometer imbalance, displayed in Fig.~\ref{fig:FransonQFC_FranG1}, is determined from the observed oscillation of the coincidence rate in this window. 

The coincidence rate in such a Franson experiment was theoretically described by Ou \cite{Ou07}: 
%
\begin{eqnarray}
R_{AB}^{\left(2\right)} \propto 1 + F\left(\tau_A-\tau_B\right) \left|\gamma_p\left(\frac{\tau_A+\tau_B}{2}\right)\right|\cdot\nonumber\\
\cdot\cos\left[\omega_A \tau_A+\omega_B \tau_B+\phi_0\right] \label{eqn:coincidencerate}
\end{eqnarray}
It oscillates in dependence of the phase between the photons (proportional to the delays $\tau_A$, $\tau_B$). The amplitude of the oscillation depends on the coherence of the pump, described by $\gamma_p(t)$, which we can set to $\gamma_p=1$ as we operate the interferometers with delays far below the pump coherence time. In our case the dominating term is
%
\begin{eqnarray}
	F(\Delta\tau) = \frac{\langle g_A(t)g^{*}_B(t+\Delta\tau) \rangle}{\langle g_A(t)g^{*}_B(t) \rangle}~,
\end{eqnarray}
using the abbreviation $g(t)=g^{(1)}(t)$. This is the normalized convolution of the first-order coherence functions of photons $A$ and $B$ of the pair; it may be attributed to the first-order coherence of the photon pair state. $F(\Delta\tau)$ has a maximum of one for equal interferometer delays ($\Delta\tau=0$) and decreases to zero for large imbalance between the delays. We calculate its expected dependence from the $g^{(1)}$ functions of the single photons derived earlier (shown in Figs.~\ref{fig:854Messungen} and \ref{fig:1310Messungen}). To compare it with the measured data we normalize the peak height to the maximum expected visibility of the apparatus, i.e.\ $F\left(0\right)=0.88\cdot0.95=0.84$. The result is shown as solid line in Fig.~\ref{fig:FransonQFC_FranG1}. 

The observed visibility reaches the maximum allowed by the apparatus and exceeds the critical value of 50~\% by more than 5 standard deviations. This is the relevant signature for the preservation of time-energy entanglement in the OFC process. The overall shape of the measured visibility also agrees with what is expected from the single-photon coherence functions, although one notes a faster fall-off for non-zero delay imbalance, i.e.\ the entanglement of the photons decoheres faster than their combined coherence. We attribute this deviation, at least partially, to the finite window size for which the coincidences are evaluated (gray area in the inset of Fig.~\ref{fig:FransonQFC_FranG1}), which does not fully exclude coincidences from the side peaks, i.e.\ from distinguishable interferometer paths. From the experimental data points we derive a pair-coherence time of $\tau_\text{pair}=12.4$~ps. 


Following the interpretation in \cite{Tan05} we define an entanglement transfer fidelity of $F_{\rm ent}=\frac{1+V_{\rm max}}{2}=91.9\pm3.9$~\% for our frequency converter, where $V_{\rm max}$ is the maximum background-corrected visibility value. Like the visibility, this fidelity reaches the limit of 92~\% set by the measurement apparatus. 
Hence we conclude that our frequency converter transfers an entangled state from NIR-NIR photon pairs to NIR-telecom photon pairs with nearly perfect fidelity. Time-energy entangled photon pairs are also suited to perform Bell tests, employing a Bell inequality for the Franson interference visibility \cite{Tit99} that predicts a boundary value $V \leq \frac{1}{\sqrt{2}} \approx 70.7\%$. Our experimentally measured maximum visibility violates this Bell inequality by 1.7 standard deviations. 

The results shown above for the two-photon interference were obtained by scanning the phase via the piezo mounted mirror M2 in the Michelson interferometer. It is worth mentioning that we observed the same visibility by scanning the phase in the Mach-Zehnder interferometer. 
This proves that the coincidence rate indeed depends on the sum of the phases of both interferometers, as predicted by equation~\ref{eqn:coincidencerate}. 

\section{Second-order Coherence}

\begin{figure}[htb]
	\centering
		\includegraphics[width=0.45\textwidth]{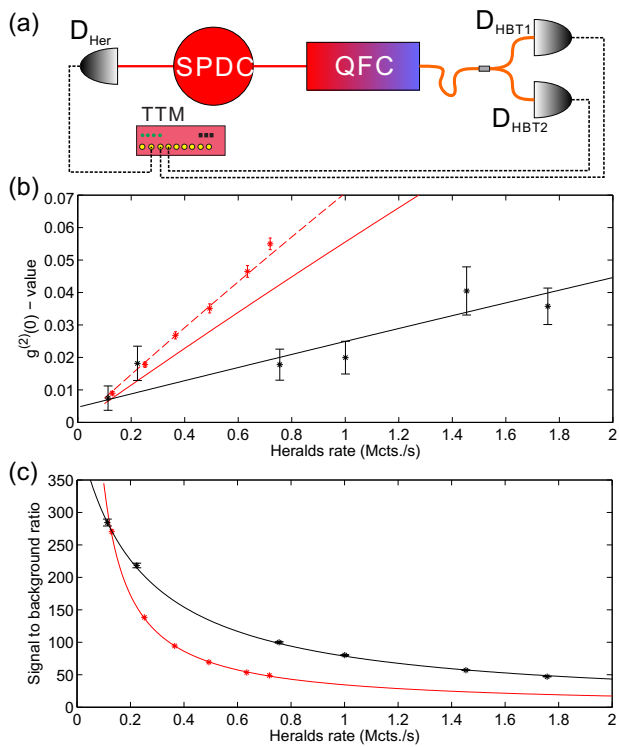}
	\caption{Photon-photon correlation measurements. a) Measurement set-up; SPDC: photon pair source, QFC: quantum frequency conversion, D$_{\rm Her}$: herald detector, D$_{\rm HBT}$: Hanbury-Brown Twiss detectors, TTM: time-tag module. For details see text. b) Measured $g^{(2)}(0)$-values vs. heralding rate. Red data points are measured at 854~nm without QFC. Black data include QFC to 1310~nm. Error bars are calculated including $\sqrt{N}$-noise of the coincidences. The solid lines result from the model based on the SBR, as described in the text. c) Signal-to-background ratio, SBR, vs. heralding rate, for a time bin of $\Delta t=1.5$~ns; the colors correspond to (b). The lines are calculations explained in the text.}
	\label{fig:HerG2Results}
\end{figure}

We also investigated the second-order coherence, i.e. the intensity correlation of the photons. A popular application of SPDC is the generation of heralded single photons. Therefore we are especially interested in the autocorrelation at zero time delay, the $g^{(2)}(0)$ value, which describes the purity of a single-photon state. In detail this means once a herald photon is detected in one mode, the partner mode is projected into a state which is a close approximation to a single photon Fock state \cite{Chr12}. The deviation from ideal single photons is due to the fact that the probability for the creation of pairs by SPDC follows a thermal distribution. The probability to generate multiple pairs at the same time increases with higher pump power. We measure the correlation with a Hanbury-Brown Twiss (HBT) set-up illustrated in Fig.~\ref{fig:HerG2Results}a. It consists of a 50/50 fiber beam splitter with both outputs connected to a SSPD. We record the time tags of the NIR herald detections and of the HBT detectors for later post-processing. To analyze the data and calculate the $g^{(2)}(0)$-values we follow the method described in \cite{Fas04}: 
we use the time tags of the heralding detector as a reference and first search the time tags of the HBT detectors for events within a 1.5~ns time window relative to the herald. The result is a new list for each HBT detector containing a 0 when there was no event for a certain herald and a 1 when a detection occurred in the particular time window. 
In the next step we correlate the two lists by sorting all pairs of detections by the number of heralds that occurred between them. The sign convention is that the distance is negative if the event on the second detector occurred before the one on the first. Finally we create a histogram of these numbers and read off $g^{(2)}(0)$ after normalization.

The results for $g^{(2)}(0)$ for varying pump power are shown in Fig.~\ref{fig:HerG2Results}b. We use the herald rate, $R_{\rm Her}$, to scale the abscissa instead of the pump power itself, after verifying their proportionality. As a reference we first measure the correlation without frequency conversion (i.e.\ with HBT detectors at 854~nm); these are the red data points. Then we include the frequency converter; the results are shown as black data points. Both measurements show the expected, approximately linear, increase. Moreover, we observe two features: first, the curve for the converted photons increases with a lower slope, and second, in the extrapolation to vanishing heralds rate the $g^{(2)}(0)$-value for the unconverted photons converges to zero while with QFC, a non-zero offset is found.

For further insight we examine the signal-to-background ratio (SBR) of the cross-correlation function of the photon pairs for each pump power value. The SBR is defined as the ratio between the coincidence rate in the bin around zero delay (Signal or true coincidences, originating from photon pairs) and the average coincidence rate in time bins with large delay (Background or accidental coincidences). We evaluate the cross-correlation between the herald photons detected with $D_{\rm Her}$ and the photons detected at one of the HBT outputs, $D_{\rm HBT1}$; Fig.~\ref{fig:HerG2Results}c shows the measured results, again for the cases without and with QFC. 

For our experiment we find an approximate expression for the SBR from the considerations detailed in the appendix: 
\begin{eqnarray}
	{\rm SBR} &=& \frac{1}{\Delta t (a R_{\rm Her} + b)} 
	\label{eqn:SBR}
\end{eqnarray}
where $\Delta t$ is the time bin size used in the experiment, and $a$ and $b$ are related to the photon loss between pair creation and detectors, and to the background detection level. 
Both SBR data sets have been fitted with Eq.~(\ref{eqn:SBR}), yielding the curves displayed in Fig.~\ref{fig:HerG2Results}c. The resulting fit parameters are consistent with the observations, see appendix. 

Starting from the model for the SBR we can also describe the results of the heralded second-order correlation measurement shown in Fig.~\ref{fig:HerG2Results}b. The $g^{(2)}(0)$-value depends on the signal to background ratio via \cite{Bec01}
\begin{equation}
	g^{(2)}(0) = 1 - \left(\frac{\rm SBR}{{\rm SBR}+1}\right)^2 
	\label{eqn:g2value}
\end{equation}
The corresponding curves are plotted as solid lines in Fig.~\ref{fig:HerG2Results}b. For the measurement with conversion (black), the prediction agrees well with the measured g$^{(2)}(0)$ values. The prediction for the case without conversion (red or gray), however, lies systematically below the measured values. We attribute this to the fact that Eq.~(\ref{eqn:g2value}) assumes a time bin size $\Delta t$ such that all coincidences are contained in the central bin around delay zero. Using $\Delta t=1.5$~ns, this is fulfilled in good approximation for the 1310~nm data, but not for the 854~nm data, because of the higher time jitter of the detectors. In fact, the discrepancy is resolved by using a 1.29 times larger time bin, as displayed by the dashed line in Fig.~\ref{fig:HerG2Results}b. 

From the model description we also conclude that the offset of the g$^{(2)}(0)$-values for the QFC measurement at vanishing SPDC pump is due to a constant background photon rate generated in the QFC process. These photons are not correlated to the SPDC signal and therefore become the dominating part of the $g^{(2)}(0)$-value at very low pump powers. At the same time, the higher SBR of the converted photons and the correspondingly lower $g^{(2)}(0)$ values are due to less uncorrelated photons that reach the detector in the 1310~nm case, partially because of better mode matching and partially because of the filtering effect of the QFC process, already observed in the first-order coherence measurements.

\section{Summary and Conclusion}

We demonstrated quantum frequency conversion of heralded single photons from 854~nm to the telecom O-band at 1310~nm with a conversion efficiency of 8~\%. We investigated the coherence properties of the photons in detail. The limited bandwidth of the conversion process results in spectral filtering, affecting first- and second-order coherence properties of the photons. In both cases we present models that describe the results well. We further used the results from first-order coherence measurements of the single photons to explain the coherence of the two-photon state, showing the dependence of the pair properties on their individual constituents. In summary, via this detailed characterization we demonstrated that the QFC process preserves the coherence properties of the photons. 

The SPDC photon pairs are intrinsically time-energy entangled. With the observation of Franson interference between converted photons and their unconverted partners, we prove that this entanglement is preserved in our frequency converter. The observed raw visibility was high enough to violate a Bell inequality.
The transfer of the time-energy entanglement from NIR-NIR photon pairs to NIR-telecom pairs was realized with a fidelity around 92~\%. Entanglement-preserving frequency conversion is an essential ingredient for long-range, fiber-based quantum networks. The high fidelity demonstrates the feasibility of including QFC in quantum information processing.

The preservation of the intensity correlation, and especially the observation of g$^{(2)}(0)\ll1$, prove the single-photon character of the converted photons. The preservation of the single-photon state during QFC has been demonstrated earlier \cite{Zas12}; here we were able to extend this observation to photons generated by SPDC. The spectral filtering effect additionally increases the single-photon purity. This opens the way for applying our system as a source of heralded single photons at telecom wavelengths.

The frequency conversion of time-energy entanglement between near infrared and telecom wavelengths shows that QFC is a versatile interface for quantum communication and quantum networks. For the source reported here the center wavelength of the SPDC photons is resonant with an atomic transition in calcium. As we are able to convert SPDC photons with our frequency converter, preserving temporal correlation, time-energy entanglement and heralded single photon state, we plan to use the converter for heralded absorption experiments \cite{Pir10, Kur14, Huw13} where we shift the heralding photon to a telecom wavelength. Furthermore, with the current setup we are able to convert single photons emitted from a single trapped ion to the telecom band. The transition at 854~nm in $^{40}$Ca$^+$ has already been employed for such atom-photon interfaces \cite{Stu12, Kur16}. In order to connect them with the telecom range, future developments of the setup will also provide polarization independent conversion, thus allowing the conversion of polarization qubits and polarization entanglement.

\section*{Funding Information}
The work was funded by the German federal ministry of science and education within the project "Q.com-Q" (contract No.~16KIS0127). J.~Brito acknowledges support by CONICYT. 


\section{Appendix}

\subsection{Derivation and evaluation of the SBR}

We consider a process where photon pairs are created, then deterministically split into two modes 1,2, and finally detected. Pair creation happens at rate $P$, the total attenuation between creation and detection is described by the factors $\eta_{1,2}$. The process is accompanied by the creation of background photon rates per mode $Q_{1,2}$ that are proportional to the pair rate (the latter being assumed proportional to the pump power). These photons do not appear in pairs; they may be produced, e.g., by Raman scattering or they may be due to imperfect mode matching from source to detector. We write $Q_{1,2} = q_{1,2}\cdot P$. Furthermore, dark counts are recorded on the two detectors at rates $W_{1,2}$. We are interested in the temporal correlation of detection events in the two modes. The correlation function is recorded in time bins of size $\Delta t$.

The detection rates (singles rates) on the two detectors are given by $S_{1,2}=\eta_{1,2}P(1+q_{1,2})+W_{1,2}$. The rate of accidental coincidences per time bin in the correlation function is therefore 
$$BG = S_1 S_2 \Delta t$$
and the total number of accidental coincidences in one bin is $BG\cdot T$, where $T$ is the total measurement time. 

The rate of true coincidences (i.e., from photon pairs) on the two detectors is given by
$$C = P\eta_1\eta_2$$
The total number of true coincidences is $C\cdot T$. We define the signal-to-background ratio as 
$${\rm SBR} = \frac{C}{BG}=\frac{C\cdot T}{BG\cdot T}$$ 
In our measurements, the dark count rate $W_1$ in mode 1, where the herald photons are detected, is negligibly small. This leads to the expression 
$${\rm SBR} = \frac{1}{\Delta t (a R_1 + b)}$$
whereby 
$$~~~~~~a = \frac{1+q_2}{\eta_1} ~~{\rm and}~~ b = \frac{1+q_1}{\eta_2}W_2~.~~~~~~{\rm (A1)}$$

Starting from a measured correlation function, one first determines the average background value from time bins sufficiently away from zero delay, then the total coincidence counts by summing up the values from the time bins around zero delay, after background subtraction. This is how the data points in Fig.~5c of the main paper have been determined. 

The two fits in this figure yield the parameters 
$$a=6.78~,~b=1.67\cdot10^6~{\rm s}^{-1}$$
for the data at 854~nm (red or gray) and 
$$a=19.1~,~b\sim0$$
for the data at 1310~nm (black). 

The difference in the values for $a$ (which ideally should be the same) are traced back to different mode matching in the two experimental runs, consistent with our observation that we measured different single rates on the detectors for the two runs. While all measurements including QFC were done together in a short period of time, the reference measurement for the second order coherence without QFC was performed later, after a realignement of the source. The value of $b$ is consistent with the measured dark count level, being approximately 2300~cts./s, on the detector for the 1310~nm photons, which is produced along with the QFC process as a constant background. 

\end{document}